\documentstyle[preprint,aps,prb]{revtex}

\begin{document}
\draft

\title{A Self--Consistent First--Principles Technique Having Linear
Scaling}

\author{E.~Hern\'{a}ndez and M.J.~Gillan \\
	Physics Department, Keele University \\
	Staffordshire ST5 5BG, United Kingdom}


\maketitle

\begin{abstract}

An algorithm for first--principles electronic structure calculations
having a computational cost which scales linearly with the system size
is presented. Our method exploits the real--space localization of the
density matrix, and in this respect it is related to the technique of
Li, Nunes and Vanderbilt. The density matrix is expressed in terms of
localized {\em support\/}~functions, and a matrix of variational
parameters, $L_{\alpha \beta}$ having a finite spatial range. The total energy
is minimized with respect to both the support functions and the
$L_{\alpha \beta}$ parameters. The method is variational, and becomes
exact as the ranges of the support functions and {\em L\/} matrix are
increased. We have tested the method on crystalline silicon systems
containing up to~216 atoms, and we discuss some of these results.

\end{abstract}


\section{Introduction}

\label{sec:intro}

There has recently been rapidly growing activity in condensed matter
simulation based on a quantum description of the electrons. The methods
being used range from simple tight--binding models~\cite{tightbinding}
to full {\em ab initio\/} techniques~\cite{abinitio}. Conventional
electronic structure methods
face severe difficulties for large systems, because the number of
computer operations generally increases as the third power of the number of
electrons. The development of methods for which the number of operations
increases only linearly with the number of electrons (linear scaling
methods) is an important target of current research. We describe
here a promising first--principles linear--scaling method, which we have
tested on silicon systems, and we present some results of these tests.

The method we propose is closely related to several recently described
techniques, particularly that of Li~{\em et al.\/}~\cite{li}. The key idea
of their method is that the electronic ground state should be determined
by variation of the total energy with respect to the density matrix,
linear--scaling being obtained by imposing a spatial cut--off on the
density matrix. This approach is already becoming widely used in the
tight--binding framework~\cite{qiu}. The method we describe is a
generalization of the approach of Li {\em et al.\/} to first principles
calculations. As we shall point out, the density--matrix approach is
related to the technique of Mauri~{\em et al.\/}
\cite{mauri:galli:car,mauri:galli},
which also has been highly successful in dynamical tight--binding
simulations. The parallel implementation of linear--scaling tight--binding
methods has also been recently reported~\cite{goedecker}. Other
linear--scaling methods less relevant to the present work have also been
described in refs.~\cite{baroni,drabold,yang,wang:teter,stechel}.

The basic principles of our method are outlined in sec.~\ref{sec:principles},
and its practical implementation is described in sec.~\ref{sec:implement}.
The results of our tests are presented in sec.~\ref{sec:tests}. Conclusions
and suggestions for future developments are given in sec.~\ref{sec:concs}.

\section{Basic Principles}

\label{sec:principles}

The first--principles method we describe is based on density functional
theory~(DFT)~\cite{abinitio}.
Within DFT, the total energy~$E_{tot}$ can be regarded as a
functional of the occupied Kohn--Sham orbitals~$\psi_i(\underline{r})$,
and the ground state
can be obtained by minimizing the total energy with respect to these
orbitals. Equivalently, the total energy can be treated as a functional of
the density matrix~$\rho(\underline{r},\underline{r}')$, defined as:
\begin{eqnarray}
\rho(\underline{r},\underline{r}') = \sum_i \psi_i(\underline{r}) \,
\psi_i^\ast(\underline{r}'),
\end{eqnarray}
where the sum goes over all occupied orbitals. The ground state can then be
obtained by minimizing~$E_{tot}$ with respect
to~$\rho(\underline{r},\underline{r}')$, subject to the
conditions that~$\rho$ is idempotent, {\em i.e.\/}:
\begin{eqnarray}
\rho(\underline{r},\underline{r}') = \int \! d \, \underline{r}'' \,
\rho(\underline{r},\underline{r}'')
           \, \rho(\underline{r}'',\underline{r}'),
\end{eqnarray}
and that the number of electrons~$N_{el}$ has the correct value, the latter
being given by:
\begin{eqnarray}
N_{el} = 2 \,\int \! d \, \underline{r} \, \rho(\underline{r},\underline{r}).
\end{eqnarray}
Whether one works in terms of~$\psi_i(\underline{r})$
or in terms of~$\rho(\underline{r},\underline{r}')$, the essence of
the calculation is to determine the occupied subspace.

If one works with the density matrix, the idempotency condition is
awkward to enforce directly, and it is more convenient to minimize subject
to the condition that all its eigenvalues lie between 0 and 1. This
corresponds exactly to the commonly used device of working with variable
occupation numbers in DFT~\cite{variable}. This can be achieved following the
strategy proposed by Li~{\em et al.\/}~\cite{li}, in which~$\rho$ is
expressed as:
\begin{eqnarray}
\rho = 3 \, \sigma \ast \sigma - 2 \, \sigma \ast \sigma \ast \sigma,
\end{eqnarray}
where~$\sigma(\underline{r},\underline{r}')$ is an auxiliary 2--point
function. Here the asterisk represents the continuum analog of matrix
multiplication, so that~{\em e.g.\/} the 2--point function
$\sigma \ast \sigma(\underline{r},\underline{r}')$ is given by:
\begin{eqnarray}
\sigma \ast \sigma(\underline{r},\underline{r}') \equiv
	\int d \, \underline{r}'' \sigma(\underline{r},\underline{r}'').
	\sigma(\underline{r}'',\underline{r}').
\end{eqnarray}
The point here is that if~$\lambda$ is an eigenvalue of~$\sigma$, then
the corresponding eigenvalue of~$\rho$ is $f(\lambda) = 3\,\lambda^2
- 2\,\lambda^3$. This transformation guarantees that if~$\sigma$ is
nearly idempotent, $\rho$ will be idempotent to an even better
approximation. The process of minimizing $E_{tot}$ has the effect of
driving the eigenvalues towards zero or unity, so that $\rho$ is
driven towards idempotency, as described in more
detail by Li~{\em et al.\/}~\cite{li}.

For practical first--principles calculations,
$\sigma(\underline{r},\underline{r}')$ must be made separable, {\em
i.e.\/} expressed in the form:
\begin{eqnarray}
\sigma(\underline{r},\underline{r}') = \sum_{\alpha, \beta}
\phi_{\alpha}(\underline{r}) \, L_{\alpha \beta} \,
\phi_{\beta}(\underline{r}'),
\end{eqnarray}
where the~$\phi_{\alpha}(\underline{r})$ will be referred to as
{\em support functions\/}. It follows
that~$\rho(\underline{r},\underline{r}')$ is also
separable:
\begin{eqnarray}
\rho(\underline{r},\underline{r}') = \sum_{\alpha, \beta}
\phi_{\alpha}(\underline{r}) \, K_{\alpha \beta} \,
\phi_{\beta}(\underline{r}'),
\end{eqnarray}
with the matrix~$K$ given by:
\begin{eqnarray}
K = 3 \, LSL - 2 \, LSLSL,
\end{eqnarray}
where $S_{\alpha \beta}$ is the overlap matrix:
\begin{eqnarray}
S_{\alpha \beta} = \int d\, \underline{r} \,
\phi_{\alpha}(\underline{r}) \, \phi_{\beta}(\underline{r}).
\end{eqnarray}
In order to turn this into a linear--scaling method, we now require
firstly that the support functions~$\phi_{\alpha}(\underline{r})$ be
non--zero only within localized spatial regions, referred to as
{\em support regions\/}, and secondly that the matrix elements
$L_{\alpha \beta}$ be non--zero only if the corresponding regions
are separated by less than a chosen cutoff distance~$R_{cut}$. It is
natural to impose these conditions, because in
general~$\rho(\underline{r},\underline{r}')$
decays to zero as the separation~$\mid\!\underline{r} -
\underline{r}'\!\mid$ goes to infinity. This implies that the
calculation will become exact as the cutoff distance and the
size of the support regions are increased.

The strategy is now to minimize the total energy both with respect
to the support functions and with respect to the~$L_{\alpha \beta}$
coefficients, subject only to the condition that the number of
electrons is held fixed at the required value. Since we are
imposing constraints on the size of the support regions and
the range of the~$L$ matrix, the calculation will be variational:
the minimum energy is an upper bound to the true ground--state
energy.

\section{Practical implementation}

\label{sec:implement}

We have implemented the above general scheme in the local density
approximation~(LDA) using the pseudopotential technique~\cite{abinitio}.
The algorithm developed in the present work
performs all calculations on a grid in real space. In this respect,
our techniques have much in common with the real--space grid
methods recently developed by Chelikowsky~{\em et
al.\/}~\cite{chelikowsky} for DFT--pseudopotential calculations. We
work with periodic boundary conditions in order to avoid surface
effects, but the technique could easily be applied with other
boundary conditions. At present, our calculations are restricted to
cubic repeating cells, and each cell is covered by a uniform
cubic grid of spacing~$\delta x$.

The support regions are chosen to be spherical with radius~$R_{reg}$,
and are centered on the atoms. Each region is associated with a
certain number~$\nu$ of support functions, where~$\nu$ is the same
for all regions. It is important to note that the total number of
support functions must be at least half the number of electrons,
but can be greater, and we exploit this freedom in the
calculations described later. Each support
function~$\phi_{\alpha}(\underline{r})$ is represented by its
values~$\phi_{\alpha}(\underline{r}_\ell)$ on the grid
points~$\underline{r}_\ell$ in its region.

We now need to evaluate the various terms in the total energy,
namely the kinetic energy~$E_K$, the electron--pseudopotential~$E_{ps}$,
the Hartree energy~$E_{H}$ and the exchange and correlation
energy~$E_{xc}$. In an exact calculation~$E_K$ would be given by:
\begin{eqnarray}
E_K = 2 \, \sum_{\alpha \beta} \int d\,\underline{r} \,
\phi_{\beta}(\underline{r}) \,
      K_{\alpha \beta} \, \left( -\frac{\hbar^2}{2 \, m} \nabla_r^2
      \right) \, \phi_{\alpha}(\underline{r}).
\end{eqnarray}
We approximate this by replacing the Laplacian by a
finite--difference approximation and the integration by a sum over
grid points. In the terminology of Chelikowsky~{\em et al.\/}
\cite{chelikowsky}, we are currently using the
second--order approximation, in which the calculation
of~$\nabla^2_r \,\phi_{\alpha}(\underline{r})$ at any grid point involves two
points
on either side in each cartesian direction. It is a simple matter to
go to higher approximations, and the computational cost of doing so is
not significant. It is important to note
that this scheme gives non--zero~$\nabla^2_r \,\phi_{\alpha}(\underline{r})$
values at
grid points on which~$\phi_{\alpha}(\underline{r})$ itself is zero, and it is
essential
to keep these values when calculating the matrix elements involved
in~$E_K$.

The energies $E_{ps}$, $E_{H}$ and $E_{xc}$ all depend on the
electron density $n(\underline{r})$, whose value at grid
point~$\underline{r}_\ell$ is:
\begin{eqnarray}
n(\underline{r}_\ell) = 2 \, \sum_{\alpha \beta}
		       \phi_{\alpha}(\underline{r}_\ell) \,
		       K_{\alpha \beta} \,
		       \phi_{\beta}(\underline{r}_\ell) .
\end{eqnarray}
The pseudopotential energy is evaluated by
multiplying~$n(\underline{r})$ by the total pseudopotential at each
grid point and summing over the grid. (For present purposes, we
are working with local pseudopotentials, although the extension to
non--local pseudopotentials is straightforward.)
The LDA exchange--correlation
energy is evaluated similarly by summing the values
$n(\underline{r}_\ell) \, \epsilon_{xc}[n(\underline{r}_\ell)]$,
where $\epsilon_{xc}(n)$ is the exchange--correlation energy per electron
at density~{\em n\/}. The Hartree energy is evaluated in reciprocal
space using the Fourier components of~$n(\underline{r}_\ell)$
obtained by discrete fast Fourier transform.

The ground state is determined by minimization of the total energy
with respect to both the support functions~$\phi_{\alpha}(\underline{r})$
and the
$L_{\alpha \beta}$ coefficients, with the electron number held
constant. We perform the minimization by the conjugate gradients
method, and for this purpose we need analytical expressions for the
derivatives~$\partial E_{tot}/\partial \phi_\alpha(\underline{r}_\ell)$ and
$\partial E_{tot}/\partial L_{\alpha \beta}$. These expressions are
straightforward to derive, as will be described in more detail in a
separate publication. The explicit formulas for these derivatives
are:
\begin{eqnarray}
\frac{\partial E_{tot}}{\partial \phi_{\alpha}(\underline{r}_\ell)} =
4 \, \sum_{\beta} [K_{\alpha \beta}
(\hat{H} \, \phi_{\beta})(\underline{r}_\ell) & + &
3 \, (LHL)_{\alpha \beta} \, \phi_{\beta}(\underline{r}_\ell) - \\
 2 \, (LSLHL & + & LHLSL)_{\alpha \beta} \,
\phi_{\beta}(\underline{r}_\ell)] \nonumber
\end{eqnarray}
and
\begin{eqnarray}
\frac{\partial E_{tot}}{\partial L_{\alpha \beta}} =
6 \, (SLH + HLS)_{\alpha \beta} - 4 \, (SLSLH + SLHLS + HLSLS)_{\alpha
\beta}.
\end{eqnarray}
Here, $(\hat{H} \, \phi_{\beta})(\underline{r}_\ell)$ denotes the function
obtained by acting with the Kohn--Sham Hamiltonian
on~$\phi_{\beta}(\underline{r})$,
evaluated at grid point~$\underline{r}_\ell$. In the matrix products
$H_{\alpha \beta}$ is the matrix element of the Kohn--Sham Hamiltonian
between support functions~$\phi_{\alpha}(\underline{r})$
and~$\phi_{\beta}(\underline{r})$. We stress that
these are exact formulas for the derivatives of the discrete
grid expressions for~$E_{tot}$. It is also worth noting that the
formula for~$\partial E_{tot}/\partial L_{\alpha \beta}$ is
identical to what would be obtained in a tight--binding formulation
with non--orthogonal basis functions.

The linear--scaling behaviour arises from the spatial localization of
the support functions, which implies that the overlap and Hamiltonian
matrices~$S_{\alpha \beta}$ and~$H_{\alpha \beta}$ vanish if the
distance between the support functions exceeds a certain cutoff.
With the cutoff we are imposing on~$L_{\alpha \beta}$, this means that
all matrices appearing in the expressions for~$E_{tot}$ and its
derivatives are sparse, and the number of non--zero elements grows
linearly with the number of atoms.

In practice, the minimization is currently performed by making a
sequence of conjugate gradients steps for the~$L_{\alpha \beta}$
coefficients, followed by a sequence of steps for the support
functions, repeating the alternation between these two types of
variation. Ultimately, more efficient procedures may prove possible.

In the tight--binding technique of Li {\em et al.\/}~\cite{li}, the
chemical potential rather than the number of electrons, $N_{el}$, was
held constant. This is inconvenient, and we have preferred to hold
$N_{el}$ fixed during the minimization. To achieve this, we project
the derivatives~$\partial E_{tot}/\partial L_{\alpha \beta}$ so that
the resulting search direction is tangential to the local surface of
constant $N_{el}$, and after each displacement of $L_{\alpha \beta}$
we make a correction to regain the correct $N_{el}$ value. In
performing this constrained minimization, there is considerable
freedom in the choice of object function, and we find that it is
convenient to minimize $E_{tot} - \mu N_{el}$, where $\mu$ is set
equal to an estimate for the chemical potential.

\section{Practical tests}

\label{sec:tests}

The total ground--state energy calculated by the above
scheme converges to the correct value as the radius $R_{reg}$ of the support
regions and the value of the spatial cutoff radius
$R_{cut}$ for the~$L_{\alpha \beta}$
coefficients are increased.  Clearly, the practical usefulness of the
method depends on the manner of this convergence. We must be able to
obtain acceptable accuracy with manageable values for $R_{reg}$ and
$R_{cut}$. The size of the region and the value of the
cutoff needed also determine the size of the system at which
linear--scaling behaviour is obtained. To test these questions, we
have performed calculations on repeating cells of perfect crystal
silicon.

The electron--core interactions are represented by the simple model
pseudopotential due to Appelbaum and Hamann~\cite{appelbaum}. This is
a local pseudopotential which is known to give a satisfactory
representation of the energetics and electronic structure of
crystalline silicon. The exchange--correlation energy is given by the
Ceperley--Alder formula~\cite{ceperley}.
We have performed tests on systems of different sizes,
using a grid spacing of 0.34~\AA. This is very similar to the spacing
typically used in pseudopotential plane--wave calculations on
silicon, and is sufficient to give reasonable accuracy~\cite{chelikowsky}.
In all cases, we have found that the conjugate gradient method converges
in a stable and fairly rapid way to the ground state. Generally,
50 steps each of~$\phi_{\alpha}(\underline{r})$ variation and~$L_{\alpha
\beta}$ variation
are more than enough to achieve convergence of~$E_{tot}$ to
within $\mbox{10}^{-4}$~eV/atom.

To examine the dependence of $E_{tot}$ on the region radius $R_{reg}$, we
have done calculations on a system of~216 atoms. For this purpose,
we have not imposed any cutoff on the~$L_{\alpha \beta}$
coefficients, but instead have determined them by exact
diagonalization of the Hamiltonian matrix after every fifth
displacement of the support functions. This is equivalent to
using an infinite cutoff for the~$L_{\alpha \beta}$ coefficients.
We stress that exact diagonalization is only used for the purpose of
this test. It is clearly not a linear scaling operation, and in
practical implementations is replaced by variation with
respect to the~$L_{\alpha \beta}$ coefficients. Our results for
$E_{tot}$ for five region sizes shown in Fig.~\ref{fig:1ehmg}
demonstrate that the
convergence of~$E_{tot}$ is very rapid, and that an accuracy of
0.05~eV/atom is reached for a region radius of 2.55~\AA. The
conventional plane--wave method needs a plane--wave cutoff of
$\approx$~12~Ry to achieve the same accuracy with commonly used
pseudopotentials for silicon. This implies that
linear--scaling behaviour for those parts of the calculation
involving the calculation of~$S_{\alpha \beta}$
and~$H_{\alpha \beta}$ matrix elements is reached for systems
of roughly 100~atoms.

The dependence of~$E_{tot}$ on the cutoff radius $R_{cut}$ for
the~$L_{\alpha \beta}$ coefficients has been studied for a
system of 216 ions. Since we have shown that a region radius of
2.21~\AA\ provides good accuracy, we have used this length to perform
this test. Here we have used conjugate gradient minimization with
respect to both~$\phi_{\alpha}(\underline{r})$ and~$L_{\alpha \beta}$.
The convergence of $E_{tot}$ with increasing $R_{cut}$ in the
$L$~matrix is illustrated in Fig.~\ref{fig:2ehmg},
where it can be seen that this
convergence is fairly rapid. The error in the total energy is less
than~1\%\ with $R_{cut} = 6$~\AA. It is worth noticing that we do not
need longer cut--offs in the {\em L\/} matrix than are needed in the
orthogonal tight--binding case~\cite{li} to obtain a similar degree
of convergence in the total energy with a given support region
radius.

\section{Discussion}

\label{sec:concs}

We have shown that our proposed first--principles linear--scaling
method is promising. The tests on c--Si show
that the total energy converges rapidly as the size of the support
regions and the cutoff radius are increased. For the time consuming
parts of the calculation involving the computation of overlap and
Hamiltonian matrix elements, the linear--scaling regime appears to
be reached with only 100~atoms. Linear scaling is reached for the
parts involving products of these sparse matrices requires larger
systems, but these operations are essentially the same
as would be needed in a tight--binding calculation.

It should be noted that there are still problems that need to be
addressed before a fully operational simulation code is
written. One such problem concerns the calculation of the
Hellmann--Feynman forces, and the possible need for Pulay
corrections~\cite{pulay}. We do not believe that this problem is particularly
severe, and we hope to address it in a later publication.
We also note that a considerable effort will be needed on code
optimization before conclusions can be drawn about the speed of the
method in practical problems.

Finally we return to the relation between our method and previous
linear--scaling schemes. Our reliance on the density matrix
techniques proposed by Li {\em et al.\/}~\cite{li}
has already been stressed. The
close relation between these techniques and the approach of
Mauri~{\em et al.\/} for tight--binding calculations has been
pointed out by Nunes and Vanderbilt~\cite{nunes}.
However, an important difference is that
in our scheme there is considerable flexibility in choosing the
number of support functions~$\nu$, whereas in the scheme of
Mauri~{\em et al.\/} it appears to be necessary to take this
number equal to half the number of electrons. We believe that
the flexibility in our scheme may be an advantage, because
we expect that increasing the number of support functions
will allow one to reduce the size of the support region, and therefore
the length of the cutoff for~$L_{\alpha \beta}$ coefficients.

The method we propose appears to be well suited to parallel
computation, and we are currently investigating possible parallel
implementations.

\section*{Acknowledgments}

Support for the postdoctoral position of E.H. is provided by
SERC grant GR/J01967. The calculations were performed partly on local
work-station facilities provided by SERC grants GR/H31783 and GR/J36266, and
partly using an allocation of time on the Fujitsu~VPX machine at
Manchester University Computer Centre under grant number
GR/J69974. We gratefully acknowledge useful discussions with
Prof.~D.~Vanderbilt, Dr.~J.~Holender and Mr.~C.~Goringe.

\figure{
Variation of the total energy with the region radius
$R_{reg}$.
\label{fig:1ehmg}
}

\figure{
Variation of the total energy with the $L_{\alpha \beta}$
matrix range, $R_{cut}$. The quantity plotted is the error in the
total energy per atom with respect to the value obtained with infinite
$R_{cut}$.
\label{fig:2ehmg}
}

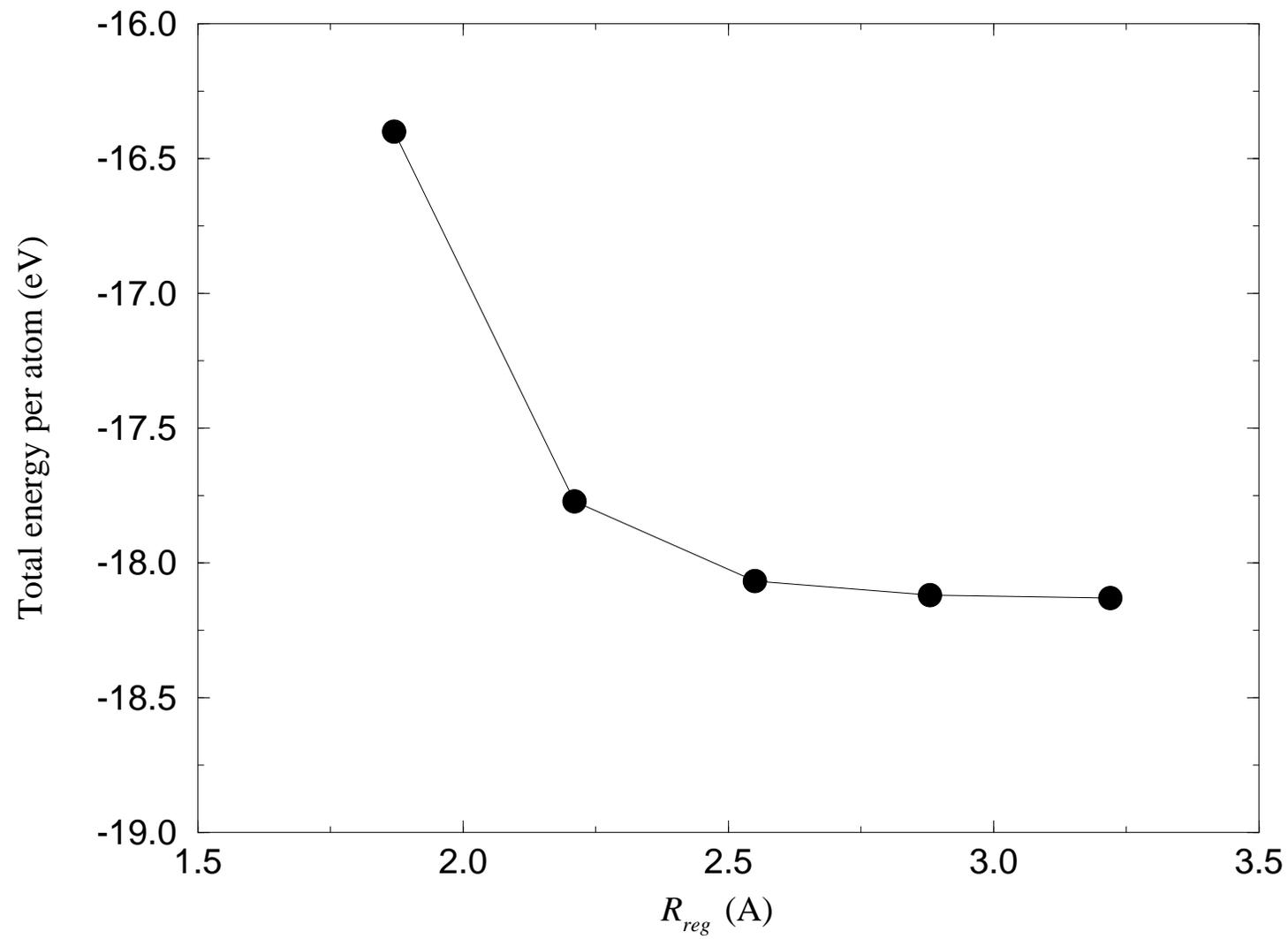
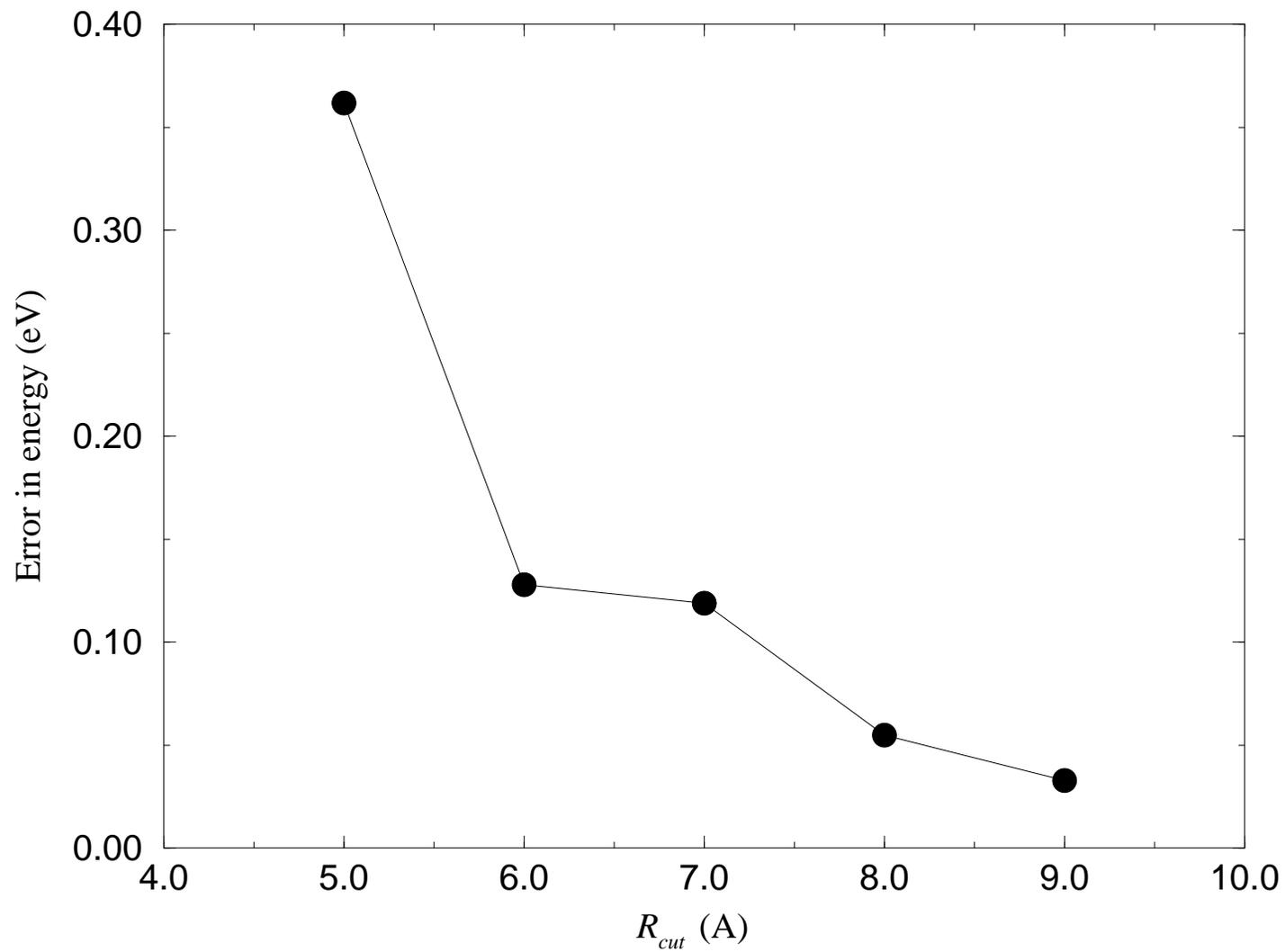
\end{document}